\documentclass[twocolumn]{openjournal}
\usepackage{lipsum}

\usepackage{color}
\definecolor{orange}{rgb}{1,0.5,0}

\usepackage[table,xcdraw]{xcolor}
\usepackage{txfonts}

\usepackage{xcolor}
\usepackage{textgreek}
\usepackage[utf8]{inputenc}
\usepackage[english]{babel}
\usepackage{graphicx}
\usepackage{placeins}

\usepackage{hyperref}
\hypersetup{
    unicode, 
    colorlinks=true,
    linkcolor=linkcolor,
    citecolor=linkcolor,
    filecolor=linkcolor,
    urlcolor=linkcolor,
}
\usepackage{color,colortbl}
\definecolor{linkcolor}{rgb}{0.0,0.3,0.5}
\usepackage{tensind}
\tensordelimiter{?}
\DeclareGraphicsExtensions{.bmp,.png,.jpg,.pdf}
\usepackage{verbatim}
\usepackage[normalem]{ulem}
\usepackage{orcidlink}
\usepackage{soul}

\graphicspath{ {./figs/} }

\begin{document}
\title{The Moon as a possible source for Earth's co-orbital bodies}

\author{R.~Sfair\orcidlink{0000-0002-4939-013X}}
\email{rafael.sfair@unesp.br}
\affiliation{UNESP - São Paulo State University, Grupo de Dinâmica Orbital e Planetologia, Av. Ariberto Pereira da Cunha, 333, Guaratinguetá, 12516-410, SP, Brazil}
\affiliation{LIRA, Observatoire de Paris, Université PSL, Sorbonne Université, Université Paris Cité, CY Cergy Paris Université, CNRS,  92190 Meudon, France}
\affiliation{Institute for Astronomy and Astrophysics, Department of Computational Physics, Eberhard Karls Universität Tübingen, Auf der Morgenstelle 10, 72076 Tübingen, Germany}

\author{L.~C.~Gomes\orcidlink{0000-0002-1191-599X}}
\email{claudio.gomes@unesp.br}
\affiliation{UNESP - São Paulo State University, Grupo de Dinâmica Orbital e Planetologia, Av. Ariberto Pereira da Cunha, 333, Guaratinguetá, 12516-410, SP, Brazil}

\author{O.~C.~Winter\orcidlink{0000-0002-4901-3289}}
\email{othon.winter@unesp.br}
\affiliation{UNESP - São Paulo State University, Grupo de Dinâmica Orbital e Planetologia, Av. Ariberto Pereira da Cunha, 333, Guaratinguetá, 12516-410, SP, Brazil}

\author{R.~A.~Moraes\orcidlink{0000-0002-4013-8878}}
\email{ricardo.moraes@unesp.br}
\affiliation{UNESP - São Paulo State University, Grupo de Dinâmica Orbital e Planetologia, Av. Ariberto Pereira da Cunha, 333, Guaratinguetá, 12516-410, SP, Brazil}
\affiliation{IFES - Federal Institute of Education, Science and Technology of Espirito Santo, Rodovia Miguel Curry Carneiro, 799, Nova Venécia, 29830-000, Espirito Santo, Brazil}

\author{G.~Borderes-Motta\orcidlink{0000-0002-4680-8414}}
\email{gabriel@asu.cas.cz}
\affiliation{Astronomical Institute of the Czech Academy of Sciences, ASU-CAS, Fričova 298, 25165 Ondřejov, Czech Republic}
\affiliation{Swedish Institute of Space Physics, IRF, Bengt Hultqvists väg 1, 981 92 Kiruna, Sweden}

\author{C.~M.~Schäfer\orcidlink{0000-0002-0341-3738}}
\email{ch.schaefer@uni-tuebingen.de}
\affiliation{Institute for Astronomy and Astrophysics, Department of Computational Physics, Eberhard Karls Universität Tübingen, Auf der Morgenstelle 10, 72076 Tübingen, Germany}

\begin{abstract}
{There is a growing number of Earth's co-orbital bodies being discovered. At 
least five of them are known to be temporarily in quasi-satellite orbits. One 
of those, 469219 Kamo'oalewa, was identified as possibly having the same 
composition as the Moon.
}
{We explore the conditions necessary for lunar ejecta to evolve into Earth's
co-orbital bodies, with particular attention to the formation of quasi-satellite
orbits. We systematically investigate the parameter space of ejection velocity
and geographic launch location across the entire lunar surface.
}
{The study employs numerical simulations of the four-body problem
(Sun-Earth-Moon-particle) with automated classification methodology for
identifying all co-orbital states. Particles are ejected from randomly
distributed points covering the entire lunar surface with velocities ranging
from 1.0 to 2.6 times the Moon's escape velocity. 
}
{Trajectories co-orbital to Earth are found to be a common outcome, with
approximately 6.15\% of all simulated particles evolving into Earth co-orbital
motion and 1.92\% specifically exhibiting quasi-satellite behavior. We identify
an optimal ejection velocity (1.2v$_{esc}$) for quasi-satellite production,
yielding over 6\% conversion efficiency at this specific velocity. The spatial
distribution of successful ejections shows a strong preference for the
equatorial regions of the trailing hemisphere. Collisions with Earth or the Moon
occur for only 4\% of the sample. %Our extended integrations reveal
%exceptionally long-lived configurations, including tadpole orbits persisting for
%10,000 years and horseshoe co-orbitals maintaining stability for 5,000 years.
}
{Our results strengthen the plausibility of lunar origin for Earth's co-orbital
bodies, including quasi-satellites like Kamo'oalewa and 2024~PT$_5$. We identify
both "prompt" and "delayed" co-orbital formation mechanisms, with a steady-state
production regime that could explain the presence of lunar-derived objects in
Earth's co-orbital regions despite the infrequent occurrence of major lunar
impacts capable of launching meter-scale fragments.
}
\end{abstract}

% Write your keywords here

\maketitle

%-------------------------------------------------------------------
\onecolumngrid
\vspace{2cm}
\twocolumngrid
\section{Introduction}
In the context of the planar circular restricted 3-body problem there are the
well known Lagrangian equilibrium points ($L_i$, $i=1, \ldots, 5$). The triangular
points, $L_4$ and $L_5$ can be stable depending on the mass ratio of the two
massive bodies \citep{Gascheau-1843}. When stable, the system
can show libration regions along the orbit of the secondary body, whose
trajectories will be under the 1:1 mean motion resonant effect.

There are several distinct solutions for this co-orbital motion, including
the tadpole and the horseshoe orbits \citep{Murray-Dermott-1999,
Namouni-Christou-Murray-1999}. Within the class of co-orbital motions there are
also the retrograde or quasi-satellite orbits. These trajectories are similar to
that of satellites, but lie far out of the Hill sphere of the secondary body,
and they are unstable in the inner Solar System.
Compound co-orbital trajectories, including one or more of these classes of
motion, with transitions between classes along the time are also possible
\citep{Wiegert-etal-1997}. In general, the quasi-satellite behaviour occurs as a
compound class of motion with the horseshoe motion. 
Theoretical studies by \citet{Christou-Georgakarakos-2021} have
demonstrated that deep Earth co-orbitals can survive for extended periods
under specific dynamical conditions.

The number of discovered bodies in co-orbital motion with respect to the Earth
has significantly increased in the last decades. At the moment, five bodies are
known to be in quasi-satellite motion \citep{Chodas-2016,
Fuente-Marcos-2016}, two in tadpole motion \citep{Connors-etal-2011,
Santana-Ros-etal-2022} and eight in horseshoe or compound horseshoe/quasi-satellite
motion \citep{Connors-etal-2002, Brasser-etal-2004,
Christou-Asher-2011}, as listed in Table \ref{table: infos}.

\begin{table}[h]
\centering
\caption{Bodies in co-orbital motion with the Earth.}
\label{table: infos} 
\begin{tabular}{c|c|c}
\hline\hline
Body & Trajectory & Diameter$^{a}$ (m) \\
\hline
\hline
 2004 GU$_9$ & Quasi-satellite & 160-360 \\
 2006 FV$_{35}$ & Quasi-satellite & 140-320 \\
2013 LX$_{28}$ & Quasi-satellite & 130-300 \\
2014 OL$_{339}$ & Quasi-satellite & 70-160 \\
Kamo’oalewa & Quasi-satellite & 30-45 \\
2010 TK$_{7}$ & Tadpole & 250-500 \\
2020 XL$_{5}$ & Tadpole & 1,100-1,260 \\
2010 SO$_{16}$ & Horseshoe & 230-480 \\
2002 AA$_{29}$ & Horseshoe & 20-30 \\
54509 YORP & Horseshoe & 150×128×93 \\
3753 Cruithne & Horseshoe & 5,000 \\
2001 GO$_{2}$ & Horseshoe-Quasi-satellite & 40-80 \\
1998 UP$_{1}$ & Horseshoe-Quasi-satellite & 210-470 \\
2009 BD & Horseshoe-Quasi-satellite & 7-15 \\
2003 YN$_{107}$ & Horseshoe-Quasi-satellite & 10-30 \\
\hline\hline
\end{tabular}
  \begin{flushleft}
  	\quad  {\footnotesize $^{a}$ Estimated size based on the magnitude.}
\end{flushleft}
\end{table}

With the significant number of known Near-Earth objects (NEO), almost 37
thousand bodies as of Feb. 2025\footnote{https://cneos.jpl.nasa.gov/stats/
totals.html}, it is expected that NEOs constitute the main source of
Earth's co-orbital bodies. Yet, the $\sim$40 meters wide object (469219)
Kamo'oalewa, currently in the most stable known quasi-satellite motion with the
Earth \citep{Fuente-Marcos-2016}, shows an L-type reflectance spectrum that
matches lunar silicate material. This points towards a possible origin of the
object as a fragment ejected from the Moon \citep{Sharkey-etal-2021}.
More recently, the $\sim$10 meters diameter object 2024~PT$_5$ was 
discovered \citep{Fuentes-Munoz-etal-2024} and has also been identified as 
potentially having lunar origin based on its spectral properties 
\citep{Kareta-etal-2025}. 
This object had a close approach with
Earth in September/November 2024 with a remarkably low velocity relative to
Earth \citep{Fuente-Marcos-Fuente-Marcos-2024}.
Additional evidence for lunar origin comes from the minimoon 2020 CD$_3$, 
which exhibited spectral properties more consistent with lunar material than 
typical near-Earth asteroids during its temporary capture phase 
\citep{Bolin-etal-2020}.

The dynamical evolution of lunar ejecta has been investigated extensively by 
\citet{gladman95}, who performed numerical simulations to 
determine the fate of material launched from the lunar surface. Their study used a 
two-stage approach, first modeling particles in the Earth-Moon system (geocentric 
phase) and then following their heliocentric evolution after escape. They found that 
approximately 20-25\% of lunar ejecta would collide with Earth over timescales of 
$\sim10^5$ years, with most impacts occurring within the first 10,000 years following a 
steep initial decline. While their work focused on determining whether lunar ejecta 
impact Earth or Moon or escape into heliocentric orbits, they did not specifically 
investigate the formation of co-orbital configurations.

Several recent studies have investigated the lunar origin hypothesis for
Earth's co-orbitals. \citet{Renu_1} performed numerical
simulations to investigate the possibility of Kamo'oalewa being ejected from
the Moon, focusing primarily on particles launched from the lunar equatorial
region. Their results supported a lunar origin for objects like Kamo'oalewa,
finding that approximately 6\% of particles launched from the lunar equator
became co-orbital with Earth within 5,000 years.

In a follow-up study, \citet{Renu_2} maintained the focus on
equatorial launch sites while examining the influence of Earth's eccentricity
on the formation of quasi-satellite orbits, concluding that Earth's
eccentricity had minimal effect on co-orbital outcomes.

\citet{Jiao_GB} took a different approach, combining SPH simulations
with N-body integrations to determine the specific ejection site of
Kamo'oalewa from the lunar surface. Their analysis suggested that Kamo'oalewa
likely originated from the Giordano Bruno crater approximately 1-10 million
years ago, finding that within 10 million years after ejection, the percentage
of particles residing in Earth co-orbital states reached at most 1\%.

The Chinese space agency's Tianwen-2 mission \citep{mission} launched on a 
Long March 3B rocket on May 28, 2025. One of its goals is to land on the asteroid
Kamo'oalewa and collect a 100~g sample that will be returned to Earth by a
capsule. The probe is expected to reach the asteroid in July 2026.
In addition to collecting material, remote sensing observations will
take place in orbit around Kamo'oalewa. The mission results will provide
critical data to test the lunar origin hypothesis and enhance our understanding
of these unique objects \citep{Venigalla-etal-2019}.

In this work, we present a comprehensive study of the conditions required
for lunar ejecta to become Earth co-orbitals, expanding previous research in
some aspects. Unlike prior studies that focused primarily on
quasi-satellite orbits \citep{Renu_1} or limited launch sites
\citep{Renu_2, Jiao_GB}, we systematically
investigate co-orbital states 
using particles ejected from the entire lunar surface with varying velocities.
Our approach employs automated classification methodology for identifying all
co-orbital states, eliminating the subjective visual inspection methods used in
some of the previous works. This comprehensive sampling across the complete lunar surface
allows us to identify not only the optimal ejection velocity for generating
Earth co-orbitals, particularly quasi-satellites, but also the geographical
distribution of favorable launch sites on the lunar surface.

The dynamical evolution is explored in the context of the restricted four-body
problem, Moon-Earth-Sun-Particle. The numerical simulations are presented in
section~\ref{simulations}. 
In section~\ref{resu} the main results are presented.
Then, in section~\ref{S-impacts}, we discuss the
kind of asteroidal impact on the surface of the Moon that would generate a
fragment the size of the known Earth's co-orbitals. Lastly, our final comments
are presented in section~\ref{S-final}.

%-------------------------------------------------------------------
\section{Numerical simulations}\label{simulations}

We integrate a system composed of the Sun, Earth, Moon, and massless particles.
The orbital elements, as well as the masses and radii of the Earth and the Moon, were
gathered from JPL Horizons ephemeris \citep{horizons}\footnote{http://ssd.jpl.nasa.gov/sbdb.cgi}
referred to epoch 2000 January 01 12:00 (JD 2451545.0).

Unlike previous studies that focused primarily on equatorial regions or limited
launch sites, our test particles were ejected from randomly distributed points
covering the entire lunar surface using the \citet{Marsaglia-1972} algorithm 
for uniform sampling on a sphere.  This uniform sampling ensures that our 
geographical distribution analysis accurately 
reflects the relative efficiency of different lunar regions for producing 
co-orbital objects, without bias toward any particular location.
The velocity vectors were primarily normal to
the surface and outward-pointing. We also tested a control sample with
non-normal velocity vectors varying randomly in azimuth while maintaining the
outward direction, confirming that launch azimuth does not significantly affect
outcome statistics.

\begin{figure}[!h]
\centering
\includegraphics[width=1.01\columnwidth]
{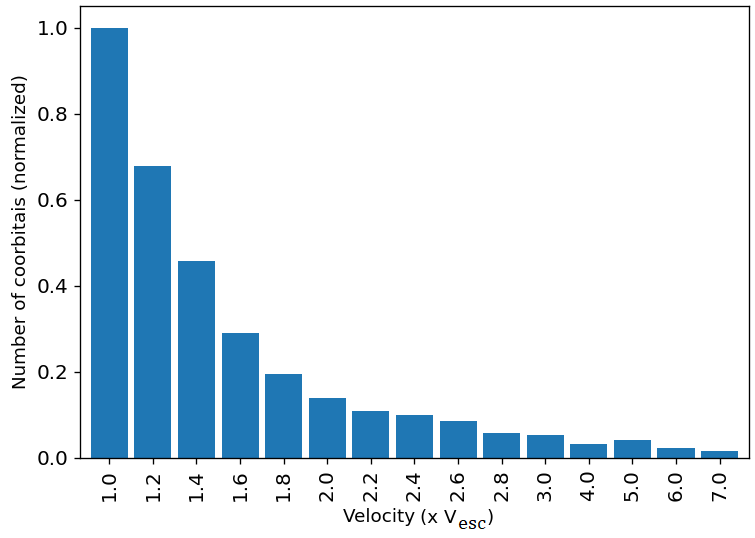}
\caption{Number of co-orbitals normalized for each simulated ejection velocity,
which ranges from 1.0 to 7.0 times the lunar escape velocity.
}
\label{QuanCoorb}
\end{figure}

Our simulations systematically explored the velocity parameter space from 1.0
up to 2.6 times the lunar escape velocity (2.4 km/s), with an ensemble of 6,000
particles for each chosen velocity, totaling 54,000 particles. 
This range of velocities was selected based on preliminary
results showing a significant drop in co-orbital production at higher ejection
speeds, as shown in Figure \ref{QuanCoorb}. To assess the independence of
results from the launch epoch, we conducted four separate simulation sets with
different initial positions of the Moon in its orbit around Earth (at quarter-period
intervals).

The system was integrated for 1,000 years with the high order integrator IAS15
available in the \textsc{Rebound} package \citep{ias15}. 
To verify integration quality, we monitored timestep evolution 
throughout sample integrations and confirmed that the adaptive timesteps 
selected by IAS15 maintained required numerical precision even during 
dynamically sensitive phases such as quasi-satellite states.
%For selected cases, we
%extended the integration time to 50,000 years with the addition of Venus, Mars,
%and Jupiter to capture long-term stability characteristics.

We implemented automated, quantitative criteria for classifying orbital states,
eliminating the subjective visual inspection methods.
Specifically, we consider a particle as co-orbital when its semi-major axis
remains between 0.99 and 1.01 au for at least 10 consecutive years. 
This time threshold excludes transient passages through 
co-orbital regions that do not represent sustained dynamical states.

The quasi-satellite
classification is applied when the difference between the mean longitude of the
particle and the mean longitude of Earth 
($\Delta\lambda = \lambda_{\textrm{particle}}-\lambda_{\textrm{Earth}}$) oscillates
around zero with an amplitude less than $7.5^\circ$ also for at least 10 years. This amplitude limit ensures 
complete capture of quasi-satellite states without missing or truncating 
capture intervals. Horseshoe
orbits are identified when $\Delta\lambda$ oscillates around $180^\circ$ 
with amplitude sufficient to
encompass the $L_4$ and $L_5$ points, while tadpole orbits librate around either $L_4$
($60^\circ$) or $L_5$ $(300^\circ$). 
Throughout each integration, we continuously monitored the mean
longitude difference between particles and Earth at every timestep,
recording entry and exit times for both quasi-satellite and general
co-orbital states according to our classification criteria. This approach
captured all co-orbital episodes and dynamical state transitions during
the 1,000-year simulations.

Our threshold-based classification approach provides a consistent, reproducible
methodology for identifying all co-orbital states (tadpole, horseshoe, and
quasi-satellite) while capturing transitions between these states, allowing for
comprehensive population statistics. Collisions are recorded when the distance
between a particle and any massive body becomes smaller than that body's radius,
at which point the particle is removed from the simulation.

%-------------------------------------------------------------------
\section{Results} \label{resu}

Our numerical simulations revealed that lunar ejecta can naturally evolve into 
various co-orbital configurations with Earth. From the complete set of 54,000 
simulated particles, approximately 6.15\% evolved into Earth co-orbital motion, 
with 1.92\% specifically exhibiting quasi-satellite behavior. The fraction of 
particles that collided with either Earth or the Moon was 4.01\% (3.46\% with 
Earth and 0.55\% with the Moon). Ejection velocity is a critical parameter 
influencing both the likelihood and stability characteristics of resulting 
co-orbital states.

\begin{figure*}[p]
    \centering
    \includegraphics[width=1.0\textwidth]
    {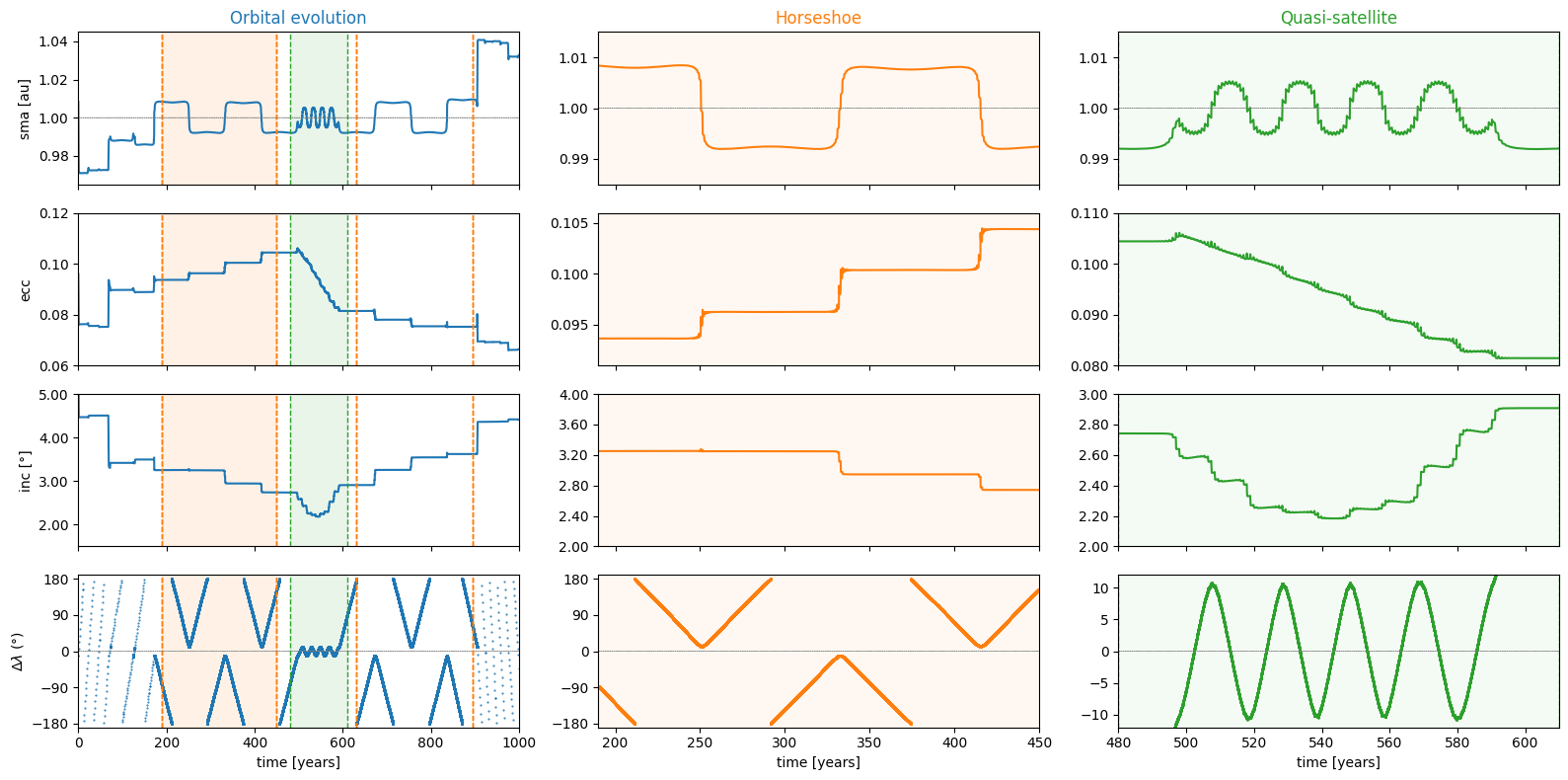}
    \caption{Example of the orbital evolution of a particle ejected from the Moon 
    with velocity 2.0 $v_{esc}$. Plots of the temporal evolution of the semi-major 
    axis, eccentricity, inclination and mean longitude with respect to the mean 
    longitude of the Earth, from top to bottom respectively. In the middle column 
    is shown a zoom of the orange region indicated in the plots of the left column.
    In the right column is shown a zoom of the green region indicated in the plots 
    of the left column.}
    \label{FigCoorColQuasi}
\end{figure*}

Fig. \ref{FigCoorColQuasi} shows a representative example of the orbital evolution 
of a particle that becomes co-orbital with the Earth. Along this trajectory, the 
horseshoe motion appears in two time periods
 ($\sim 200$ yr and $\sim 600$ yr). 
Between the two stages in horseshoe motion, the particle experiences a 
quasi-satellite motion for a period of about one hundred years. The temporal evolution 
of the orbital elements (semi-major axis, eccentricity and inclination) clearly 
shows a chaotic behaviour. However, during co-orbital phases, the orbital 
elements exhibit significantly more regular behavior, with notably reduced 
variations in semi-major axis and more constrained eccentricity excursions. This 
regularity is particularly evident during the stages of horseshoe (orange plots) 
and quasi-satellite (green plots) motions.

\begin{figure}[h]
    \centering
    \includegraphics[width=1.\columnwidth]
    {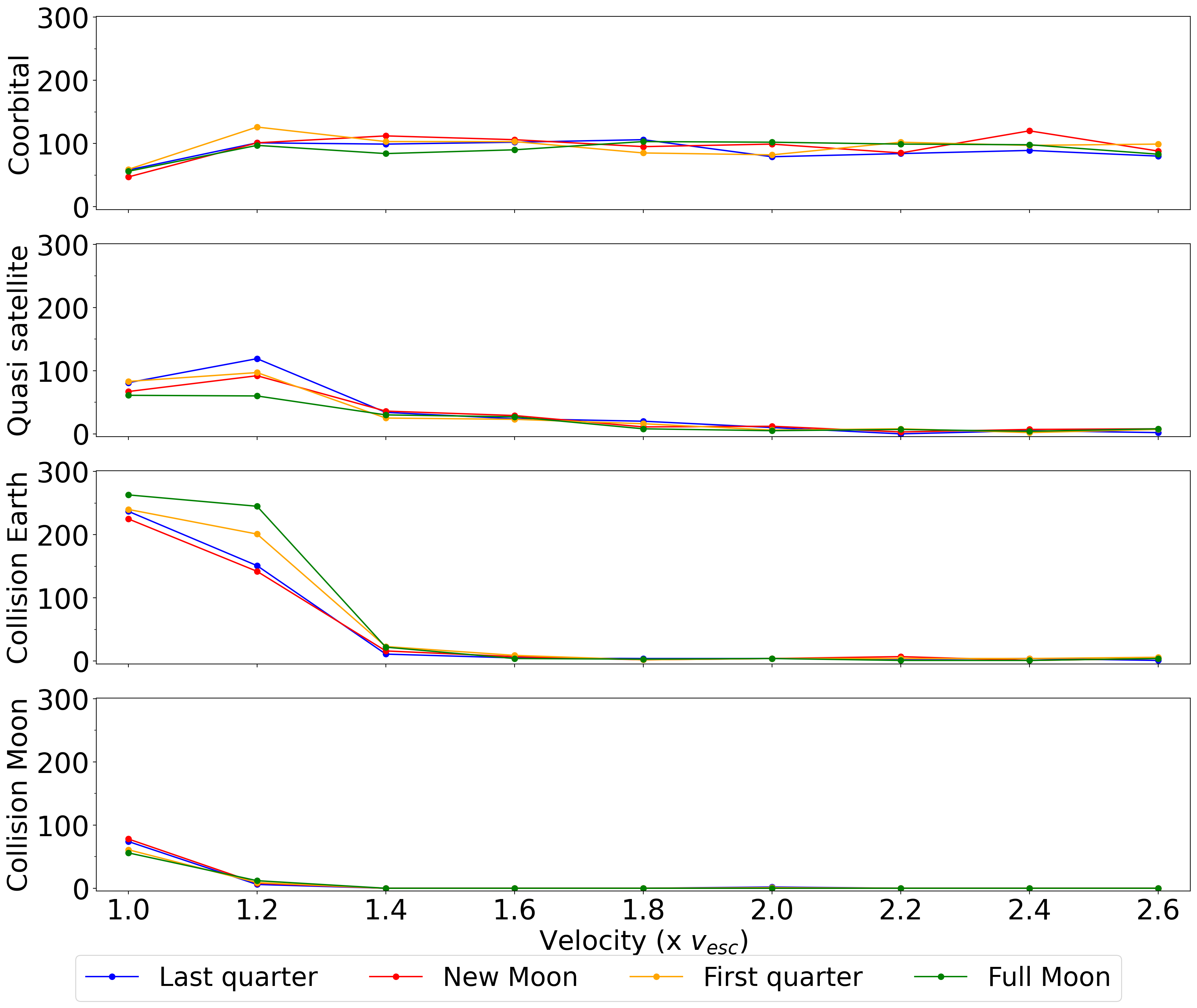}
    \caption{Simulation outcomes as a function of ejection velocity normalized 
    to the lunar escape velocity ($v_{esc}$). Plots show the number of particles 
    that became co-orbital, quasi-satellite, collided with Earth, and collided with 
    the Moon (from top to bottom). Colors represent four sets of simulations with 
    different initial positions of the Moon in its orbit around Earth, at 
    quarter-period intervals.}
    \label{outcome}
\end{figure}

The main results of the full set of simulations are presented in Fig. \ref{outcome}.
Each color represents one of the four simulation sets with different initial 
positions of the Moon on its orbit around Earth (at quarter-period intervals).
In general, the results show that there is no significant difference between them, 
confirming that the launch epoch with respect to the Sun-Earth-Moon configuration 
does not appreciably affect co-orbital production rates, neither the collisions with 
the Earth or the Moon.

The outcomes in Fig. \ref{outcome} are presented as a function of the ejection 
velocity. From top to bottom, the plots show the number of bodies that became 
co-orbital, quasi-satellite, collided with the Earth and collided with the Moon, 
respectively. Collision frequencies with both the Moon and Earth decline 
rapidly for ejection velocities exceeding 1.4$v_{esc}$. The quasi-satellite 
formation exhibits a distinct non-monotonic trend, beginning at moderate levels for 
1.0$v_{esc}$, reaching maximum production at 1.2$v_{esc}$ (exceeding 6\% of all 
particles at this velocity), before gradually decreasing at higher velocities. 
This velocity-dependent pattern suggests an optimal ejection window for 
quasi-satellite formation. The overall co-orbital count shows minimum 
production at the lowest ejection velocity, then stabilizes across the intermediate 
velocity range before gradually declining at the highest velocities tested.

Our findings both complement and extend recent studies on the lunar origin of
Earth co-orbitals. While \citet{Renu_1} observed that approximately
6\% of particles launched from the lunar equator became co-orbital with Earth
within 5,000 years, our comprehensive survey covering the entire lunar surface
yields a comparable 6.15\% co-orbital production rate, validating their results
while providing a more complete spatial distribution. However, our finding that
1.92\% of all particles exhibit quasi-satellite behavior is significantly higher
than previously reported values. \citet{Jiao_GB} found that within 10 million
years after ejection from Giordano Bruno crater, the percentage of particles in
Earth co-orbital states reached at most 1\%, with only rare cases becoming
long-lived quasi-satellites. This discrepancy likely stems from our finding of an
optimal ejection velocity (1.2$v_{esc}$) for quasi-satellite production, which
yields over 6\% conversion efficiency at this specific velocity.

Unlike \citet{Renu_2}, who focused on the influence of Earth's
eccentricity on co-orbital formation while maintaining equatorial launch sites,
our work systematically explores the parameter space of ejection velocity and
geographic location. While they concluded that Earth's eccentricity had minimal
effect on co-orbital outcomes, our results demonstrate that both velocity and
launch location significantly impact co-orbital production and stability.
Additionally, our automated classification methodology captures the full range of
co-orbital states (tadpole, horseshoe, and quasi-satellite), whereas previous
studies primarily focused on horseshoe-quasi-satellite transitions. This broader
classification approach enabled our discovery of exceptionally long-lived cases
with stability timeframes an order of magnitude longer than previously
identified.

\begin{figure}[h]
    \centering
    \includegraphics[width=1.01\columnwidth]
    {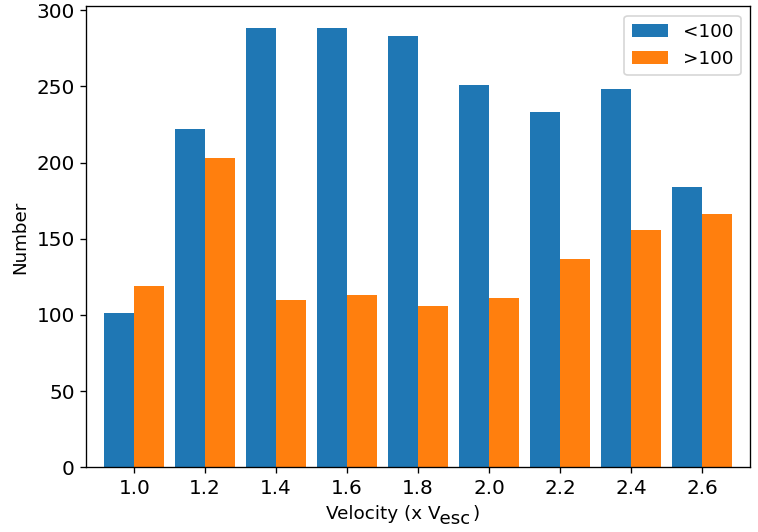}
    \caption{
Distribution of co-orbital objects as a function of initial ejection
velocity, categorized by by residence time. Blue bars represent objects with
co-orbital durations less than 100 years, while orange bars indicate those
exceeding 100 years. Each object is counted only once, regardless of multiple
co-orbital entries.
}
\label{hist_ve}
\end{figure}

To further understand how ejection velocity influences co-orbital stability, 
we analyzed the residence time distribution of objects in these dynamical states. 
Fig.~\ref{hist_ve} shows the distribution of co-orbital objects as a function
of initial ejection velocity, grouped by their residence time in co-orbital
states. We classify objects into two categories: short-lived ($<$100 years) and
long-lived ($>$100 years). Importantly, we count each particle only once in this
analysis, regardless of how many times it may enter a co-orbital configuration
during the simulation, to focus on the possibility of particles to
achieve co-orbital states. 

\begin{figure}[h]
    \centering
    \includegraphics[width=1.01\columnwidth]
    {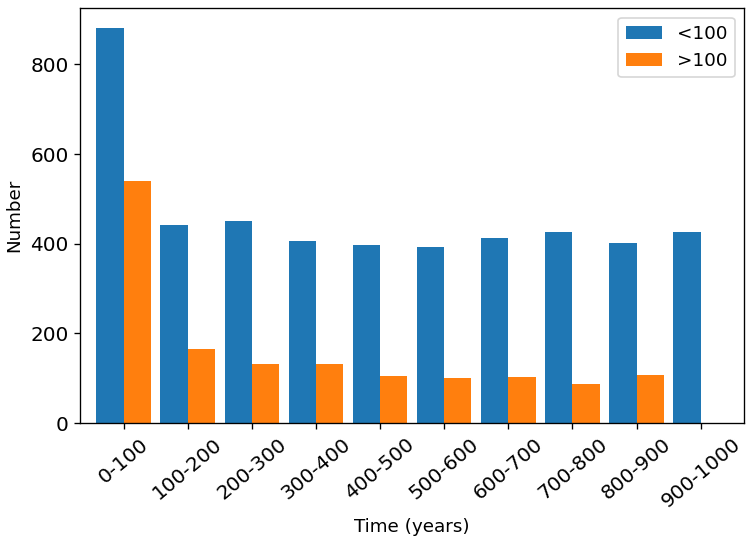}
    \caption{
    Temporal distribution of co-orbital entries throughout the 1,000-year
    simulation, binned in 100-year intervals. Blue bars represent entries into
    short-lived ($<$100 years) co-orbital states, while orange bars indicate
    entries into long-lived ($>$100 years) states. Unlike Fig.~\ref{hist_ve}, this
    histogram counts each entry event separately, allowing multiple counts per
    particle. 
    }
    \label{hist_ti}
\end{figure}

Some particles experience multiple co-orbital episodes during the 
integration, including cases where the same object exhibits both short-lived 
($<100$ years) and long-lived ($>100$ years) co-orbital periods. For the 
classification in Fig.~\ref{hist_ve}, we assign each particle to the category 
corresponding to its longest co-orbital duration to avoid double counting. 
However, to preserve information about multiple co-orbital entries and the 
full range of co-orbital behaviors, Fig.~\ref{hist_ti} records every 
individual entry event into co-orbital states, allowing us to capture the 
complete temporal evolution of co-orbital formation throughout the simulation.

The distribution reveals a strong velocity dependence in co-orbital formation.
For long-lived objects (orange bars), a clear peak occurs at
$1.2v_{\mathrm{esc}}$, followed by a gradual decrease at higher velocities.
In contrast, short-lived co-orbitals (blue bars) exhibit their maximum at
$1.6v_{\mathrm{esc}}$. A particularly interesting feature appears at the
minimum velocity of $1.0v_{\mathrm{esc}}$, where we observe an inversion in
the typical pattern - long-lived co-orbitals exceed short-lived ones, suggesting
that particles barely escaping lunar gravity preferentially establish stable
configurations when they achieve co-orbital states.

This velocity-dependent behavior provides insight into conditions for
generating Earth's co-orbital bodies. The pronounced peak at
$1.2v_{\mathrm{esc}}$ for long-lived co-orbitals coincides with our
observation that this velocity also produces the highest proportion of
quasi-satellite trajectories (exceeding 6\% of all outcomes at this velocity).
The correlation between ejection velocity and orbital longevity demonstrates
how initial conditions strongly influence the dynamical character of resulting
co-orbital states.

Fig.~\ref{hist_ti} presents the temporal distribution of co-orbital entries 
throughout our 1,000-year simulation. Unlike Fig.~\ref{hist_ve}, which counts 
each particle only once, this analysis records every individual entry event into 
a co-orbital state, allowing multiple counts per object. The histogram reveals 
two distinct phases in co-orbital formation from lunar ejecta: an initial 
peak during the first 100 years followed by a relatively uniform production 
rate that persists for the remainder of the simulation.

The pronounced peak in the first time bin (0--100 years) contains approximately 
1,000 short-lived entry events and 170 long-lived ones, significantly exceeding 
the subsequent time intervals. This initial surge represents particles that 
quickly transition into co-orbital states following ejection from the lunar 
surface. After this early phase, entry rates stabilize at approximately 500 
events per 100-year interval for short-lived states and about 75 events per 
interval for long-lived configurations.

This transition to a steady-state production regime suggests a fundamental 
separation between ``prompt'' and ``delayed'' co-orbital formation mechanisms. 
The prompt mechanism captures particles that directly enter co-orbital 
configurations through relatively straightforward dynamical pathways. In 
contrast, the delayed mechanism involves more complex orbital evolution, where 
particles undergo sufficient perturbations in the Earth-Moon-Sun system before 
eventually achieving co-orbital states.

The near-constant rate of co-orbital entries following the 
initial surge suggests the establishment of a kind of steady-state  in co-orbital production.
This implies that lunar material can 
continuously replenish Earth's co-orbital population through stochastic 
processes, even centuries after the initial impact event.
This persistent production mechanism could explain the presence of lunar-derived 
objects like Kamo'oalewa and 2024 PT$_5$ in Earth's co-orbital regions despite 
their inherently temporary orbital configurations. Considering that major lunar 
impacts capable of launching meter-scale fragments occur infrequently, the 
steady-state mechanism provides a pathway for lunar material to enter co-orbital 
states across geological timescales.

The full statistical analysis of our 54,000 simulated particles yielded 
6.15\% of trajectories evolving into Earth co-orbital motion, with 
1.92\% specifically exhibiting quasi-satellite behavior. Collisions 
occurred in only 4.01\% of cases (3.46\% with Earth and 0.55\% 
with the Moon). We found that ejection velocity strongly influences co-orbital outcomes, with  
particles ejected at 1.2$v_{\mathrm{esc}}$ produced the highest proportion 
of quasi-satellites, exceeding 6\% at this specific velocity. This 
optimal ejection window represents a significant refinement compared to 
earlier studies that reported lower co-orbital conversion efficiencies 
\citep{Jiao_GB}. 

Our Earth collision rate differs from the 20-25\% reported
by \citet{gladman95} over $10^5$ years due to our focus on velocities optimized for
co-orbital formation. Their finding that most collisions occur within the first
10,000 years aligns with our observed temporal distribution. The differences in
outcome statistics primarily reflect our exploration of higher velocity regimes and
focus on co-orbital formation rather than direct Earth impact.

Our integration timeframes revealed exceptional orbital stability in 
certain cases, far exceeding previous estimates. The 1,000-year simulations 
identified quasi-satellites and horseshoe co-orbitals maintaining their 
respective states throughout the entire integration period. 
%More 
%significantly, our extended 50,000-year integrations discovered extremely 
%long-lived configurations: tadpole orbits persisting for 10,000 years and 
%horseshoe co-orbitals maintaining stability for 5,000 years. These 
%longevity values substantially exceed the stability timeframes reported by 
%\citet{Renu_1}, who primarily observed persistence on 
%timescales of hundreds of years. 
%The significantly longer orbital stability we observed might be attributed to 
%our approach of sampling the complete lunar surface rather than equatorial 
%regions alone and systematically exploring ejection velocities, 
%which revealed the critical $1.2~v_{\mathrm{esc}}$ 
%threshold for generating stable co-orbital configurations.

%-------------------------------------------------------------------
\section{Lunar origin of co-orbitals} \label{S-impacts}
% Schäfer: analytical approach
We have evidence for ejected materials from the Moon based on both
empirical meteorite data \citep{marvin-1983} and numerical simulations
\citep{head-2002}. Images taken by the Lunar Orbiter III and V suggested
that {40}-{100}~{m} boulders ejected from lunar craters are
not unusual. However, the ejection speeds of these boulders are typically
very low \citep{BART2010337}. Using the classic Melosh theory for
spallation \citep{melosh1985}, a {13.5}~km diameter impactor
is necessary to generate escaping ejecta blocks with {50}~m in
size.

By mapping and statistical analysis of six lunar secondary crater fields,
\cite{singer-2020} find an estimation for the largest fragment size of the
ejecta at escape velocity. They find for Copernicus and Kepler, with crater
diameters of {93}~km and {31}~km, respectively,
largest fragment sizes of {50}~m and {30}~m, which fit
the size range of Kamo'oalewa.

In \citet{Jiao_GB} they attempt to determine the location of the
Kamo'oalewa ejection from the lunar surface with numerical simulations
and SPH (smoothed particle hydrodynamics). \citet{Jiao_GB} suggest
that the ejection occurred in the impact that generated the Giordano Bruno
crater due to the age of the crater. The Giordano Bruno crater is 22 km in
diameter and is located approximately 36$^\circ$ North and 103$^\circ$
East \citep{Morota2009}. The results of our simulations agree with the
possibility of the ejection from this location.

\begin{figure*}[!ht]
\centering
\includegraphics[width=2.\columnwidth]{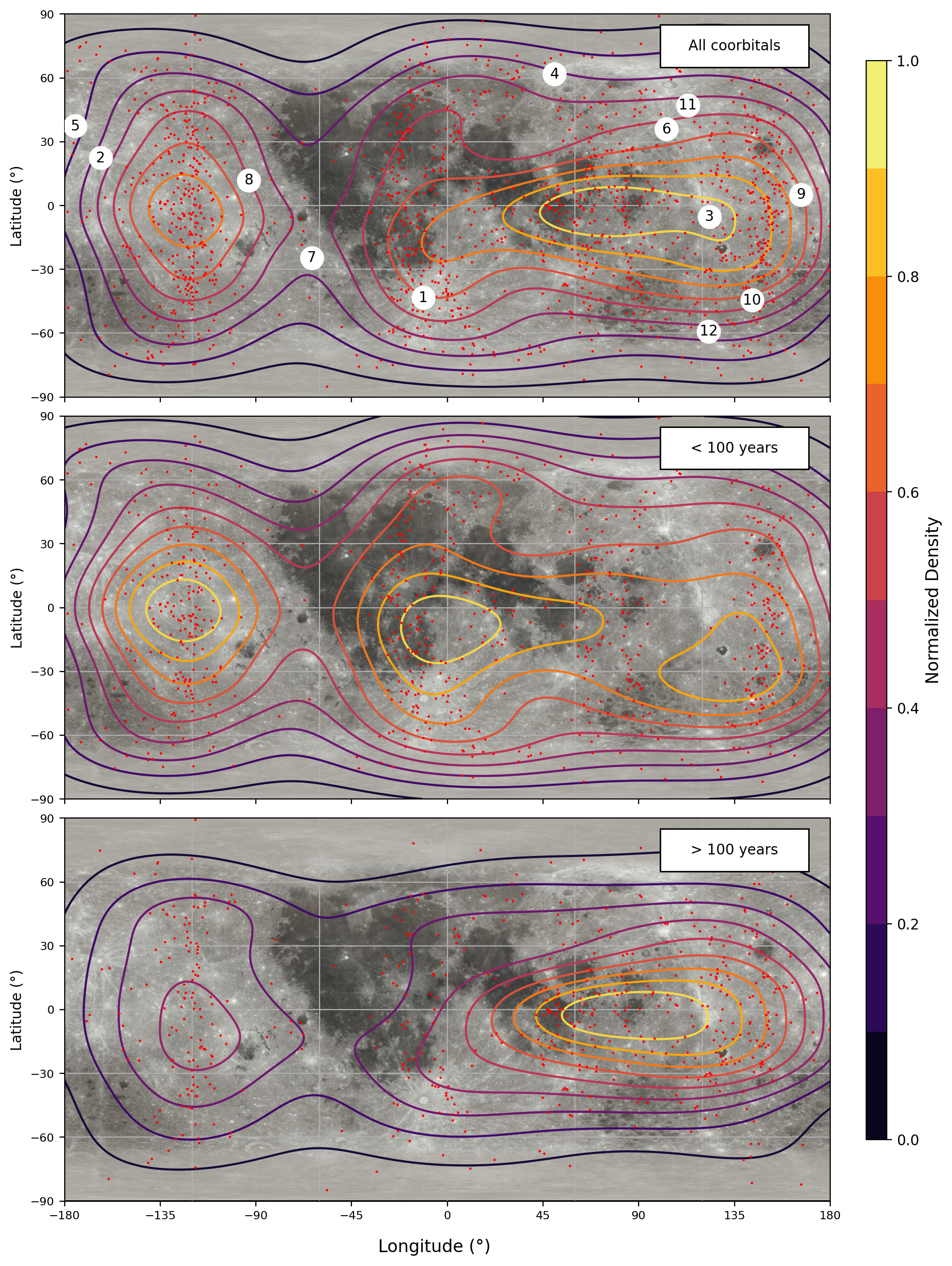}
\caption{Distribution of initial conditions on the lunar surface
that result in Earth co-orbital objects. The background grayscale
image represents the lunar surface, with contour lines showing the
normalized density of successful co-orbital objects. Red dots
indicate specific launch locations. {Top panel:} All co-orbital
objects. Numbered labels (1-12) indicate craters from Table \ref{table: crater}.
{Middle panel:} Short-term co-orbital objects with
durations less than 100 years. {Bottom panel:} Long-term
co-orbital objects with durations exceeding 100 years. Co-orbital objects
with lifetimes shorter than \textbf{10 years} were not considered. \vspace{0.2cm}}
\label{fig:coorbital_map}
\end{figure*}

Fig.~\ref{fig:coorbital_map} shows the geographical distribution of initial 
conditions on the lunar surface that lead to Earth co-orbital objects. 
The map uses a lunar coordinate system where longitude $0^{\circ}$ points 
toward Earth, longitude $-90^{\circ}$ represents the leading side (the 
direction of lunar orbital motion), and longitude $90^{\circ}$ corresponds 
to the trailing side. 
The geographical distribution analysis properly accounts for the
spherical lunar surface geometry, with density calculations normalized per
unit area to ensure accurate representation despite the rectangular
projection used for visualization.

The density distribution reveals a latitudinal 
dependence, with a higher concentration of ejections that resulted in 
co-orbital motion occurring near the equatorial regions (latitude 
$\approx 0^{\circ}$). The map was generated using particle data at all 
simulated velocities and was aligned to match the four different 
Earth-Moon-Sun geometries. This equatorial preference is observed 
in the trailing side of the Moon (longitude $\approx 90^{\circ}$), where 
the normalized density reaches its maximum values. A secondary 
concentration appears in the leading hemisphere (longitude 
$\approx -90^{\circ}$), although with lower density values. This 
indicates that equatorial ejections, especially those from the trailing 
side, tend to be more efficient at producing Earth co-orbital objects.

When comparing the middle and bottom panels, we observe differences 
in the spatial distribution between short-lived ($<100$ years) 
and long-lived ($>100$ years) co-orbital objects. Long-term 
co-orbitals show a more concentrated distribution, primarily originating 
from the trailing hemisphere with an equatorial preference. In contrast, 
short-term co-orbitals exhibit a broader distribution across the lunar 
surface, including contributions from mid-latitude regions. 
These differences suggest that the longevity of co-orbital states may 
depend on the ejection location. This concentration pattern is consistent 
with the distribution of quasi-satellites and collisions, in agreement 
with \cite{Renu_1}.

The lunar craters in Table \ref{table: crater} are taken from \cite{craters},
which analyzes and quantifies the age of lunar craters with diameter $\geq 10$ km and less than 1 Gyr.
For this study, a cut was made in the named lunar craters with less than 150 Myr and diameter $\geq 15$ km.

Examining the numbered craters in the top panel, we find that several 
coincide with regions of higher density in the co-orbital distribution. 
Craters located near the equatorial zone of the trailing 
hemisphere (particularly craters 3 and 9) align with areas showing 
enhanced production of co-orbital objects. Similarly, crater 8 in the 
leading hemisphere corresponds to the secondary density peak observed at 
longitude $\approx -135^{\circ}$. The geographical correlation 
between these specific craters and regions favorable for generating 
co-orbital objects suggests that ejection events from these locations 
could have contributed to the population of Earth co-orbitals. We can 
observe that the equatorial band opposite to the lunar movement is 
favorable for generating co-orbitals, contrariwise the polar regions.

Our results suggest that detailed laboratory analysis of lunar
meteorites or returned samples could potentially constrain their launch
locations on the lunar surface. The strong geographical preferences we
observe, particularly the enhanced production from equatorial regions of
the trailing hemisphere, indicate that compositional and chronological
data from samples could be compared against known crater distributions
and ages in these favorable regions to narrow down potential source
locations.

\begin{table}[h]
\centering
\caption{{List of named craters younger than 150 Myr and larger than 15 km in diameter, 
order by diameter. Adapted from \citet{craters}}.}
\label{table: crater} 
\begin{tabular}{c|c|c|c|c}
\hline\hline
n  & Name           & Lat (deg) & Long (deg) & D(km) \\
\hline\hline
1  & Tycho          & 43.3 S    & 11.22 W   & 85      \\
2  & Jackson        & 22.4 N    & 163.1 W   & 71      \\
3  & Necho          & 5.25 S    & 123.24 E  & 37      \\
4  & Thales         & 61.8 N    & 50.3 E    & 32      \\
5  & Moore F        & 37.4 N    & 175.0 W   & 24      \\
6  & Giordano Bruno & 35.9 N    & 102.89 E  & 22      \\
7  & Byrgius A      & 24.5 S    & 63.7 W    & 19      \\
8  & Sundman V      & 11.9 N    & 93.5 W    & 18      \\
9  & Mandel'stam F  & 5.2 N     & 166.2 E   & 16      \\
10 & Ryder          & 44.5 S    & 143.2 E   & 16      \\
11 & Rayet Y        & 47.2 N    & 113.0 E   & 15      \\
12 & Fechner T      & 59.1 S    & 122.9 E   & 15      \\
\hline\hline
\end{tabular}
\vspace{0.5cm}
\end{table}

%\vskip 50pt
\clearpage
%-------------------------------------------------------------------
\section{Final remarks} \label{S-final}

This work explored the conditions under which lunar ejecta can evolve into
Earth's co-orbital bodies through comprehensive numerical simulations of the
four-body problem (Sun-Earth-Moon-particle). By systematically sampling the
entire lunar surface and a range of ejection velocities, we identified key
factors governing the production of Earth co-orbitals, with special attention
to quasi-satellite configurations.

The numerical simulations demonstrate that lunar ejecta can evolve into Earth
co-orbital configurations with limited frequency. Ejection velocity
strongly influences co-orbital formation, with particles launched at
1.2$v_{esc}$ yielding the highest quasi-satellite production efficiency.
Geographic analysis reveals preferential co-orbital formation from equatorial
regions of the trailing hemisphere, consistent with theoretical expectations
for lunar material capture into Earth's co-orbital zones.
These findings support the lunar origin hypothesis for Earth's
co-orbital bodies, including quasi-satellites like Kamo'oalewa and 2024 PT$_5$.
The comprehensive parameter space exploration reveals both ``prompt" and
``delayed" co-orbital formation mechanisms. The delayed mechanism exhibits
steady-state production rates over centuries, providing a pathway for
continuous replenishment of Earth's co-orbital population between major
lunar impact events.
The automated classification methodology developed here provides
quantitative criteria for identifying all co-orbital states (tadpole,
horseshoe, and quasi-satellite), establishing a reproducible framework for
population statistics that captures transitions between dynamical states.

Our results indicate that lunar ejecta may remain in Earth's vicinity
or impact our planet, with implications for lunar resource extraction
activities. Excavation of lunar regolith requires mechanical forces
\citep{Wilkinson2007} that could accelerate particles beyond the 2.38 km/s
escape velocity \citep{Just2020}, particularly in the low-gravity lunar
environment.
International space law frameworks currently lack specific provisions
for debris liability from extraterrestrial mining operations
\citep{Hubbard2024}. The identified preferential ejection regions suggest
that mining activities in the lunar trailing hemisphere equatorial zones
present elevated risks for generating Earth-directed debris. Implementation
of debris mitigation protocols will be essential for safe lunar resource
extraction.

Future investigations incorporating additional planetary perturbations and
extended integration times may reveal longer-lived co-orbital configurations.
The relationship between ejection parameters and orbital stability merits
further study to constrain the origins of specific Earth co-orbital objects.
Comparative analysis of the delayed capture mechanism with resonant
phenomena in other solar system contexts may yield insights into general
capture dynamics in multi-body systems.

\vspace{10cm}
%-------------------------------------------------------------------
\section*{Data availability}
The particle outcome data from the numerical simulations presented in this
work are publicly available at \url{https://doi.org/10.5281/zenodo.16857659}.

%-------------------------------------------------------------------
\section*{Acknowledgments}
 We are grateful to the anonymous referee whose insightful suggestions significantly improved this work.
 
 This study was financed in part by 
 the Brazilian Federal Agency for Support and Evaluation of Graduate Education (CAPES), 
 in the scope of the Program CAPES-PrInt, process number 88887.310463/2018-00, 
 International Cooperation Project number 3266, 
 Fundação de Amparo à Pesquisa do Estado de São Paulo (FAPESP) - Proc. 2022/11783-5, 
 Conselho Nacional de Desenvolvimento Científico e Tecnológico (CNPq) - Proc. 316991/2023-6 and 307400/2025-5. 
 RS and CS acknowledge support by the DFG German Research Foundation (project 446102036).

%-------------------------------------------------------------------
\newpage
\bibliographystyle{apsrev4-1}

% You should give the same name for your .bbl as your main .tex
% since it is a requirement for posting on ArXiv.
\bibliography{paper_oja}

\end{document}